\documentclass[runningheads]{llncs}
\usepackage{graphicx}
\usepackage[svgnames]{xcolor}
\usepackage{float}
\usepackage{cancel}
\usepackage{tikz}
\usepackage{amsmath}
\usetikzlibrary{arrows,decorations,backgrounds,shapes, snakes}
\usetikzlibrary{positioning}
\usetikzlibrary{matrix} 
\usepackage[normalem]{ulem}               % to striketrhourhg text
\newcommand\redout{\bgroup\markoverwith
	{\textcolor{red}{\rule[0.5ex]{2pt}{0.8pt}}}\ULon}
\begin{document}

%Definitions: equations
\newcommand{\beq}{\begin{equation}}
\newcommand{\eeq}{\end{equation}}
\newcommand{\lb}{\label}
\newcommand{\ph}{\phantom}
\newcommand{\beqar}{\begin{eqnarray}}
\newcommand{\eeqar}{\end{eqnarray}}
\newcommand{\barr}{\begin{array}}
\newcommand{\earr}{\end{array}}
\newcommand{\jump}{\parallel}
\newcommand{\Ehat}{\hat{E}}
\newcommand{\That}{\hat{\bf T}}
\newcommand{\Ahat}{\hat{A}}
\newcommand{\chat}{\hat{c}}
\newcommand{\shat}{\hat{s}}
\newcommand{\khat}{\hat{k}}
\newcommand{\muhat}{\hat{\mu}}
\newcommand{\mc}{M^{\scriptscriptstyle C}}
\newcommand{\mei}{M^{\scriptscriptstyle M,EI}}
\newcommand{\mec}{M^{\scriptscriptstyle M,EC}}
\newcommand{\hbeta}{{\hat{\beta}}}
\newcommand{\rec}[2]{\left( #1 #2 \ds{\frac{1}{#1}}\right)}
\newcommand{\rep}[2]{\left( {#1}^2 #2 \ds{\frac{1}{{#1}^2}}\right)}
\newcommand{\derp}[2]{\ds{\frac {\partial #1}{\partial #2}}}
\newcommand{\derpn}[3]{\ds{\frac {\partial^{#3}#1}{\partial #2^{#3}}}}
\newcommand{\dert}[2]{\ds{\frac {d #1}{d #2}}}
\newcommand{\dertn}[3]{\ds{\frac {d^{#3} #1}{d #2^{#3}}}}

\title{Configurational forces on elastic structures}
%
%\titlerunning{Abbreviated paper title}
% If the paper title is too long for the running head, you can set
% an abbreviated paper title here
%
\author{Davide Bigoni\inst{1}\orcidID{0000-0001-5423-6033} \and
Federico Bosi\inst{2}\orcidID{0000-0002-3638-5307} 
\and
Francesco Dal Corso\inst{1}\orcidID{0000-0001-9045-8418}
\and
Diego Misseroni\inst{1}\orcidID{0000-0002-7375-671X}
}
\authorrunning{D. Bigoni et al.}
% First names are abbreviated in the running head.
% If there are more than two authors, 'et al.' is used.
%
\institute{DICAM, University of Trento, via Mesiano 77, Trento, Italy 
\and
Department of Mechanical Engineering, University College London, Torrington Place, London, WC1E 7JE, UK}
\maketitle              % typeset the header of the contribution
\begin{abstract}
The discovery of configurational forces acting on elastic structures and 
its initial applications are reviewed. 
Configurational forces are related to the possibility that an elastic structure can 
change its configuration, thus inducing a variation in the potential energy. 
This concept has already led to several applications (the elastica arm scale, the dripping of an elastic rod, and the torsional actuator), has been shown to strongly affect  stability, and to be related to limbless locomotion. It is believed that these results will open a new research territory in mechanics. 

\keywords{Eshelbian mechanics \and Configurational force  \and Elastica \and Structural stability \and Locomotion \and Deployable structures.
}
\end{abstract}

\section{Introduction}

The concept of configurational force was introduced by Eshelby \cite{eshelby1,eshelby2,eshelby3,eshelby4} to describe the tendency of defects to 
move inside solids. In particular, it is postulated that the movement occurs in a way that the total potential energy of the mechanical system decreases, until eventually it reaches a minimum, 
corresponding to a configuration where these forces vanish. 
The defects can be massless, such as voids, cracks, vacancies, and 
dislocations, or can possess a mass, such as inclusions. Within this framework, 
it may be easier to figure out the movement of a massless defect, e.g. a dislocation in a crystal lattice, than a stiff inclusion
in a material. However, the latter can be identified as a portion of material belonging to
a phase different from that of the surrounding material, so that the boundary of the inclusion will grow or shrink to minimize the energy.

The configurational force, also called `Eshelbian', or \lq material', or \lq driving', or \lq non-Newtonian', is defined as the negative gradient of 
the total potential energy $V$ of a body with respect to a parameter $\kappa$ determining the configuration
of the defect, namely, $-\partial V(\kappa)/\partial \kappa$.

Well-known examples of configurational forces are the Peach-Koehler interaction between dislocations, the crack-extension force in fracture mechanics, or the material force developing on a phase boundary in a solid under loading. These forces are the central concept in 
Eshelbian mechanics, a well-consolidated and famous theory  (see, for instance, the monographs by Gurtin \cite{gurtin}, Kienzler and Herrmann \cite{kienzler}, 
and Maugin \cite{maugin1,maugin2}, and the journal
special issues by Dascalu et al. \cite{dascalu}, and Bigoni and Deseri \cite{bigoni-deseri}).

Despite the broad diffusion of Eshelbian mechanics, examples of configurational forces developing and acting
on elastic structures were unknown before the work by 
Davide Bigoni, Federico Bosi, Francesco Dal Corso and Diego Misseroni, the authors of this 
article, who opened this field in \cite{2bodabomi}. 
The key concept relies on introducing a particular constraint, which allows a change in configuration of the elastic system. When this change leads to a variation of the 
total potential energy of the structure, an Eshelby-like force develops. A simple constraint permitting the development of a configurational force is the sliding sleeve, which constrains a 
portion of an elastic rod inserted into it. This constraint prevents rotation and transverse displacement, but allows for
axial frictionless movement. The portion of the rod inside the sliding sleeve can be considered as a \lq defect', which might change 
its position, so inducing a variation in the total potential energy and generating a related configurational force. 

\begin{figure}[ht]
  \begin{center}
 \includegraphics[width=\textwidth]{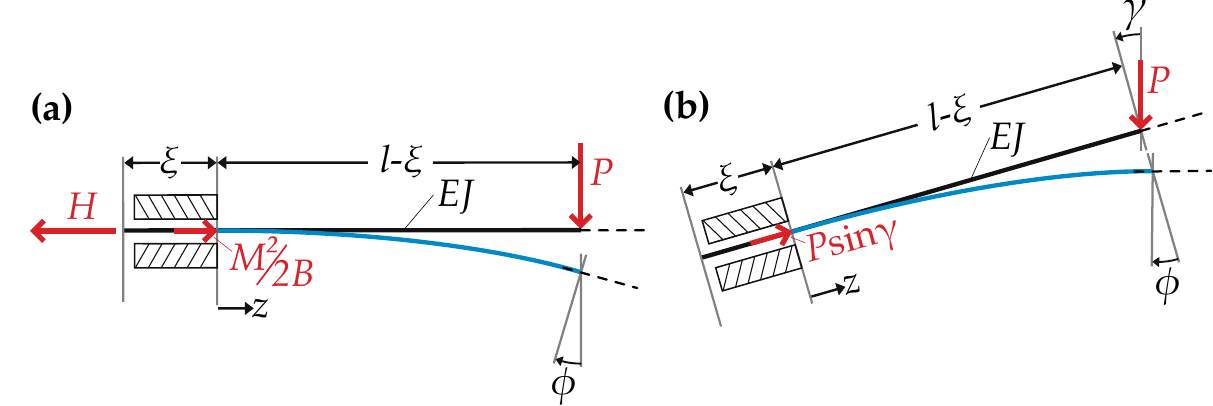}
\caption{\footnotesize (a) A simple elastic structure showing the existence of a configurational force. The key to understanding why this force develops is the presence of the sliding sleeve on the left end of an elastic rod of total length $l$, constraining a length $\xi$. 
The elastic rod is subjected to a dead vertical load $P$ on its right end and to an axial dead force $H$ applied at its left end. 
The presence of the Eshelby-like force $P^2(l-\xi)^2 /(2B) = P\phi$ (where $B=EJ$ is the rod's bending stiffness and $\phi$ is the rotation of the loaded end of the rod) changes the force $H$ at equilibrium, not null whenever the rod is bent. (b) In the absence of the axial force, $H=0$, the equilibrium of the inclined sliding sleeve (the load has a transverse component equal to $P\cos\gamma$) is obtained only when $\gamma = \phi$. When $\phi < \gamma$  
the rod slips inside the sliding sleeve, while, when 
$\phi > \gamma$, the rod is ejected outside the sliding sleeve.
}
\label{rega_1}
 \end{center}
\end{figure}

The simplest configurational force is developed at the end of a sliding sleeve constraining an initially straight elastic rod, subject to a transverse load $P$, Fig. \ref{rega_1}(a). 
%%%%%%%%%%%%%%%%%%%%%%%%%%%%%%%%%%%%%%%%%%%%%%%%%%%%%%%%
%%%%%%%%%%%%%%%%%%%%%%%%%%%%%%%%%%%%%%%%%%%%%%%%%%%%%%%%%
This structure was analyzed in a fully non-linear regime and the configuration 
force was demonstrated using two different approaches, namely, a variational technique 
and an asymptotic method, the latter based on the definition of an imperfection in the constraint. 
Finally, the structure has been 
realized and instrumented, so that the configurational force has been validated through experimental measures.

\begin{figure}[ht]
  \begin{center}
 \includegraphics[width=1\textwidth]{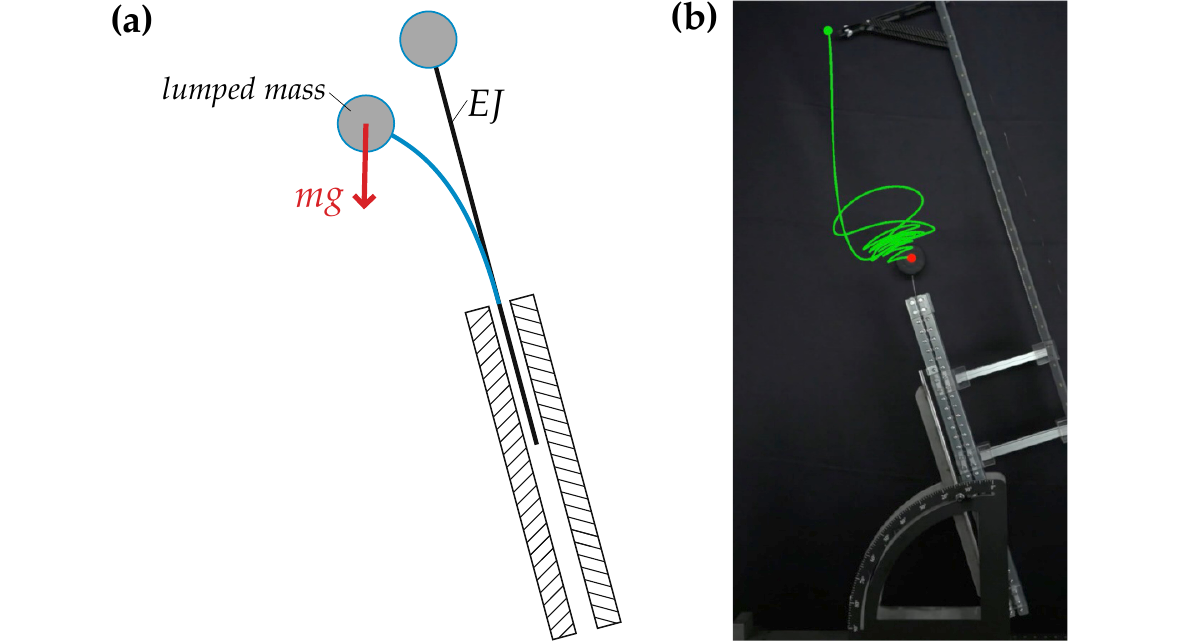}
\caption{\footnotesize (a) Schematic of the experiment showing the effects of a configurational force on the dynamics of an elastic rod equipped with a lumped mass. 
(b) Snapshot of an experiment showing the trajectory of the lumped mass (green line) and evidencing an oscillatory motion during which the elastic rod slips (\textit{into} and \textit{out from}  the sliding sleeve) 
because of the action of a configurational force.
}
\label{dynamic}
 \end{center}
\end{figure}

The presence of the configurational force was also theoretically derived and experimentally verified within a  
 dynamic framework. More specifically, Armanini et al. \cite{1armanini} have shown that the configurational force deeply influences the dynamics of an elastic rod with a lumped mass at one end and constrained with a frictionless sliding sleeve at the other, as shown in Fig.\,\ref{dynamic}. Moreover, it has also been proven that configurational forces acting at the ends of an elastic rod can strongly influence band gaps in the wave dispersion diagram 
and introduce a nonlinear coupling between longitudinal and transverse displacements \cite{10datamomobi}.

%%%%%%%%%%%%%%%%%%%%%%%%%%%%%%%%%%%%%%%%%%%%%%%%%%%%%%%%
%%%%%%%%%%%%%%%%%%%%%%%%%%%%%%%%%%%%%%%%%%%%%%%%%%%%%%%%%

In this article, the effects of configurational forces on one-dimensional elastic structures  are reviewed, together 
with applications to soft devices (the elastica arm scale, the dripping of an elastic rod,
and the torsional actuator), implications on stability (the penetration on an elastic blade), 
and connections with limbless locomotion (the snaking rod). 

It is suggested that the concept of configurational force may lead to  a new and exciting research field, {\it configurational structural mechanics}, connected to the development of 
devices to be employed in soft robotics, actuation, deployable structures, and deformable sensors.

\section{Configurational force in a simple elastic structure}

With reference to Fig. \ref{rega_1}(a), an elastic rod (straight in its unloaded configuration) of uniform bending stiffness $EJ$ and total length 
$l$ is considered, constrained by a (frictionless) 
siding sleeve and subjected to a transverse and an axial force, respectively $P$ and $H$, applied at the two ends. Under the small rotation hypothesis, the generic configuration of the system is defined by the transverse displacement field $v(z)$, where $z$ is the curvilinear coordinate measured from the sliding sleeve exit.
Moreover, the configuration is also described by the configurational parameter $\xi$, defining the position of the sliding sleeve exit along the rod, and therefore the length $l-\xi$ of the rod undergoing flexural deformations.
The total potential energy $V$ can be written as
\beq
\lb{potenzialazzo}
V(v, \xi) = \frac{EJ}{2} \int_0^{l-\xi} \left(v''\right)^2 \, dz - P \int_0^{l-\xi} v' \, dz
-H \xi,
\eeq
where a dash means derivation with respect to the 
coordinate  $z$. The transverse displacement field is subjected to the following kinematic conditions at the sliding sleeve exit
\beq\label{bbcc}
v(0)=v'(0)=0.
\eeq

Following a variational approach, the 
transverse displacement $v$ and the length $\xi$ are subjected to variations $\tilde{v}$ and $\tilde{\xi}$ through a small parameter $\epsilon$ as 
\beq
v \longrightarrow v + \epsilon \, \tilde{v},
\,\,\,\,\,\,
\xi \longrightarrow \xi + \epsilon \, \tilde{\xi}.
\eeq
By considering the kinematic boundary conditions (\ref{bbcc}), the constraints on  the variation $\tilde{v}$ in the transverse displacement field at the sliding sleeve exit are
\beq\label{bbcc2}
\tilde{v}(0)=\tilde{v}'(0)=0. 
\eeq
Annihilation of the first variation (in $\epsilon$) of the total potential energy $V$ is
\beq
\lb{variazz}
EJ  \, \int_0^{l-\xi} v''\tilde{v}''\, dz - \frac{EJ \left[v''(l-\xi)\right]^2}{2}
 \tilde{\xi}
-
P\, \int_0^{l-\xi} \tilde{v}'\, dz+ \left[P v'(l-\xi) 
- H \right] \tilde{\xi}= 0.
\eeq

Considering eqn (\ref{bbcc2}), the first integral in equation (\ref{variazz}) can be evaluated through integration by parts as
\beq
 \,\int_0^{l-\xi} v''\tilde{v}''\, dz =
\,  v''(l-\xi)\,\tilde{v}'(l-\xi)-
\, \int_0^{l-\xi} v'''\tilde{v}'\, dz .
\eeq
and therefore the first variation  (\ref{variazz}) reduces to  
\beq
\begin{array}{lll}
\lb{variazz2}
\displaystyle
-  \int_0^{l-\xi} \left( EJ\, v''' + P \right)\,\tilde{v}'\, dz 
 +EJ\,  v''(l-\xi)\tilde{v}'(l-\xi)\\[4mm]
  - \left[H+ \dfrac{EJ \left[v''(l-\xi)\right]^2}{2}-P v'(l-\xi) \right]\tilde{\xi} = 0.
\end{array}
\eeq

Taking into account eqn (\ref{bbcc2}), the integral in equation (\ref{variazz2}) can be further integrated  by parts as 
\beq
\lb{variazz4}
-  \int_0^{l-\xi} \left( EJ\, v''' + P \right)\,\tilde{v}'\, dz 
= 
\left[EJ\,v'''(l-\xi)+P\right]\tilde{v}(l-\xi) - EJ 
\int_0^{l-\xi}  v''''\,\tilde{v}'\, dz ,
\eeq
so that annihilation of the first variation under arbitrary  variations $\tilde{v}$ yields to the governing equation  of the Euler elastica (in its linearized version)
\beq\label{elastichina}
 v''''(z) = 0,\qquad z\in[0,l-\xi]
\eeq
complemented, in addition to the kinematic boundary conditions (\ref{bbcc}), by the static boundary conditions
\beq\label{bbcc3}
v'''(l-\xi)=-\frac{P}{EJ}, \,\,\,\,\mbox{ and }\,\,\,\, v''(l-\xi) = 0,
\eeq
representing the shear and moment  conditions at the end of the rod,  $z=l-\xi$. 

The arbitrariness of $\tilde{\xi}$ allows concluding from eqn (\ref{variazz2}) that axial equilibrium does not always correspond to $H=0$ as it would follow by ignoring the configurational force. More specifically, $H$ corresponds to  a non-null value when the rod is bent and is expressed by
\beq\label{eqsliding}
H = P \phi,
\eeq
where $\phi$ is the rotation at the loaded end, $\phi=v'(l-\xi)$.
Through integration of the linearized elastica, eqn (\ref{elastichina}), under  the boundary conditions (\ref{bbcc}) and (\ref{bbcc3}), the loaded end rotation $\phi$ results
\beq
\phi = \frac{P(l-\xi)^2}{2EJ},
\eeq
and therefore the equilibrium of the system evidences the presence of a {\it configurational force}
\beq
H=\frac{P^2(l-\xi)^2}{2EJ}. 
\eeq

%%%%%%%%%%%%%%%%%%%%%%%%%%%%%%%%%%%%%%%%%%%%%%%%%%%%%%%%
%\begin{figure}[ht]
%  \begin{center}
%\includegraphics[width= 8 cm]{figures/rega_2.eps}
%\caption{\footnotesize The equilibrium of the inclined sliding sleeve (the load has a transverse component equal to $P\cos\gamma$) is obtained when $\gamma = \phi$, where $\phi$ is the rotation of the loaded end of the rod. When $\phi < \gamma$  
%the rod slips inside the sliding sleeve, while, when 
%$\phi > \gamma$, the rod is ejected outside the sliding sleeve.
%}
%\label{rega_2}
% \end{center}
%\end{figure}
%%%%%%%%%%%%%%%%%%%%%%%%%%%%%%%%%%%%%%%%%%%%%%%%%%%%%%%%%

If the elastic rod is inclined at an angle $\gamma$ with respect to the loading $P$, so that its transverse and axial components are respectively $P \cos \gamma$ and $P \sin \gamma$,  the equilibrium equation (\ref{eqsliding}) would change to
\beq\label{eqslidinginc}
H+P\sin \gamma=P \phi \cos \gamma,
\eeq
where the rotation $\phi$ of the loaded end of the rod is now
\beq
\phi = \frac{P\sin \gamma(l-\xi)^2}{2EJ}.
\eeq
From  eqn (\ref{eqslidinginc}), the equilibrium in the case of null force $H=0$ (Fig. \ref{rega_1}(b)) implies  the following geometric condition
\beq
\lb{seghina}
\phi=\tan \gamma,
\eeq
which, assuming the angle $\gamma$ to be sufficiently small to allow the retainment of the linear term, leads to
\beq
\lb{angolini}
\phi = \gamma,
\eeq
implying an orthogonality condition between the applied load and the rod's tangent.

The geometric condition   (\ref{angolini}) to attain equilibrium has been proven to hold  also in the case of large rotations and large angles  $\gamma$ and shows that  \cite{2bodabomi}   
\begin{itemize}
\item when $\phi = \gamma$ the rod is at equilibrium;
\item when $\phi < \gamma$ the rod slips inside the sliding sleeve;
\item when $\phi > \gamma$ the rod is ejected outside the sliding sleeve.
\end{itemize}

\section{Elastica arm scale}

\begin{figure}[!th]
  \begin{center}
 \includegraphics[width=\textwidth]{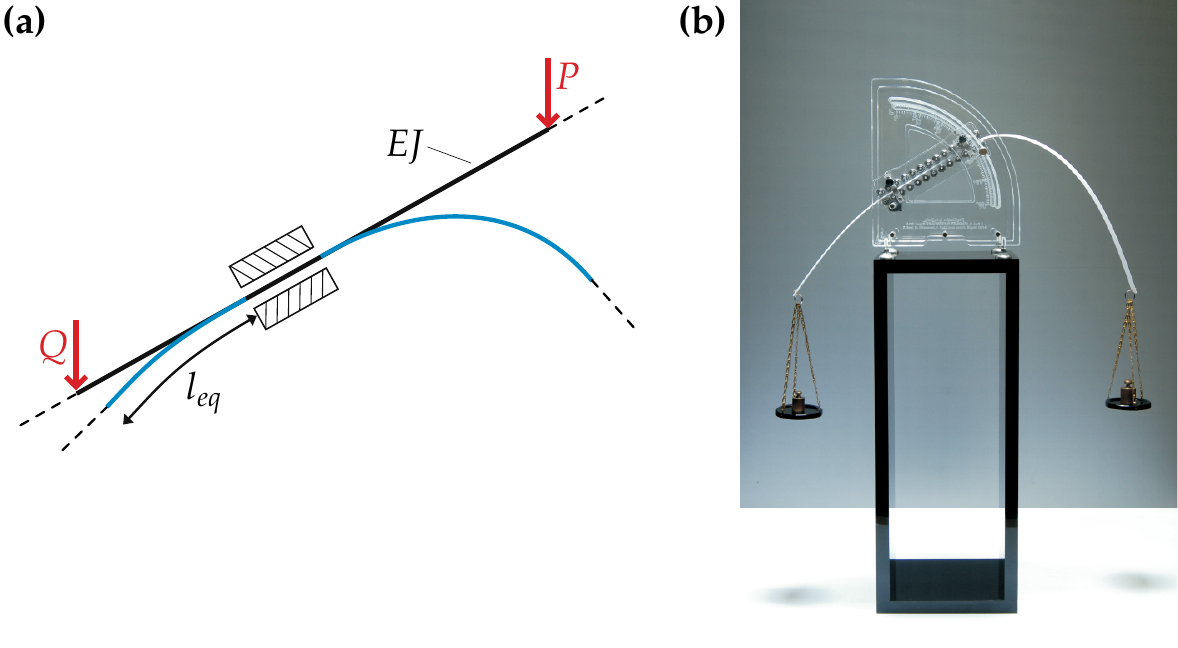}
\caption{\footnotesize (a) Schematic of the elastica arm scale based on an inclined sliding sleeve. The equilibrium is attained in the sliding sleeve direction through the contributions of the dead loads $P$ and $Q$, and the two configurational forces arising from the elastic bending of the rod. (b) A prototype of the elastica arm scale realized by D. Misseroni and donated by the authors to the \lq Museo della Bilancia' in Campogalliano (Italy).
}
\label{rega_3}
 \end{center}
\end{figure}

With reference to Fig. \ref{rega_3}(a), the concept underlying the elastica arm scale is the result of nonlinear equilibrium kinematics of a rod 
constrained centrally by a sliding sleeve, which 
induces two  
configurational forces, with outward direction from each sliding sleeve exit and opposite to each other. Therefore, the deflection of the arms becomes necessary for equilibrium, which would be impossible for a rigid system. 
The rigid arms of ordinary scales are replaced by a flexible elastic lamina, so that the rod can reach a unique equilibrium configuration when two
vertical dead loads are applied at its ends. This configuration has been analytically solved by Bosi et al. \cite{8bodamibi}, showing that the knowledge of a given  load value $P$ (at the right end) and the measure of the equilibrium length $l_{eq}$ allows for  the identification of an unknown load $Q$ (at the left end). 

\begin{figure}[!h]
  \begin{center}
 \includegraphics[width=\textwidth]{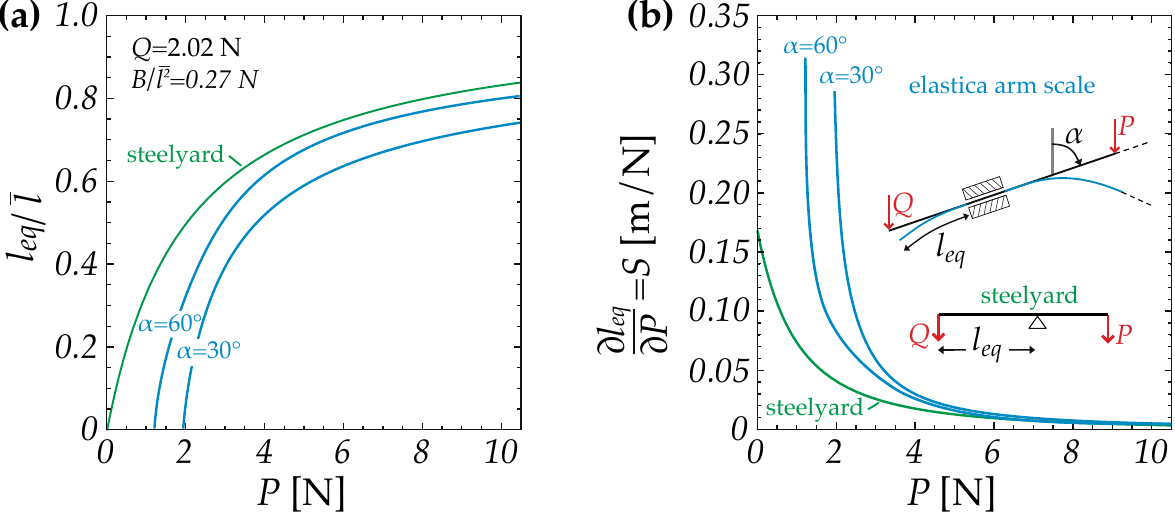}
\caption{\footnotesize Comparison between a steelyard scale and the elastica arm scale, the latter inclined at two different angles $\alpha$. (a) Equilibrium length $l_{eq}$ for different applied loads $P$. (b) Sensitivity  $S$ as a function of the applied load $P$.
}
\label{bilancia}
 \end{center}
\end{figure}

In a sense, the elastica arm scale is realized through the combination of two different mechanical principles, each one underlying the operation of a specific family of classical scales: (i.) equilibrium (on which the steelyard scale is based) and (ii.) deformation (on which the spring scale is based).  A  comparison between the elastica arm scale and the ordinary steelyard scale is shown in Fig. \ref{bilancia}. The equilibrium length $l_{eq}$ and the sensitivity (denoted by $S$) for different applied loads $P$ are shown in Fig. \ref{bilancia}(a) and (b), respectively. It follows that at small given load $P$, the sensitivity $S$ of the elastica arm scale can become superior to the 
traditional device. 

Prototypes of the elastica arm scale have been designed, realized, and tested at the Instabilities Lab of the University of Trento. The prototype shown in 
Fig. \ref{rega_3}(b) has been realized by D. Misseroni and donated by the authors to the \lq Museo della Bilancia' in Campogalliano (Italy).

\newpage
The mechanics of the elastica arm scale has been further investigated by O'Reilly \cite{or1,or2}, who considered the material momentum balance law for rods. 
%%%%%%%%%%%%%%%%%%%%%%%%%%%%%%%%%%%%%%%%%%%%%%%%%%%%%%%%
%%%%%%%%%%%%%%%%%%%%%%%%%%%%%%%%%%%%%%%%%%%%%%%%%%%%%%%%%

\vspace{-0.8cm}

%%%%%%%%%%%%%%%%%%%%%%%%%%%%%%%%%%%%%%%%%%%%%%%%%%%%%%%%
%%%%%%%%%%%%%%%%%%%%%%%%%%%%%%%%%%%%%%%%%%%%%%%%%%%%%%%%%

\section{Dripping of an elastic rod}

The application of configurational forces presented in this section addresses the so-called \lq self-encapsulation' problem, 
in which an elastic rod is loaded with a transverse force applied at midspan (between two constraints kept at a fixed distance), with the purpose of reaching a closed deformation loop, 
thus encapsulating a finite region. Self-encapsulation has connections to micro- or nano-fabrication technologies for deployable
structures used in sensors. In this field of application, self-assembly can be achieved
through magnetic forces \cite{vella}, while a self-folding spherical 
structure has been invented \cite{shim} and a dynamic self-encapsulation
technique for a thin plate and a rod has already been pointed out \cite{antk,rivetti}. 
In the former case, only a reduction in the volume of a sphere is achieved, while in the latter self-encapsulation is obtained as a result of both dynamic effects and capillary forces, which are related to the presence of
a liquid droplet attached to the rod. 

\begin{figure}[H]
  \begin{center}
 \includegraphics[width=\textwidth]{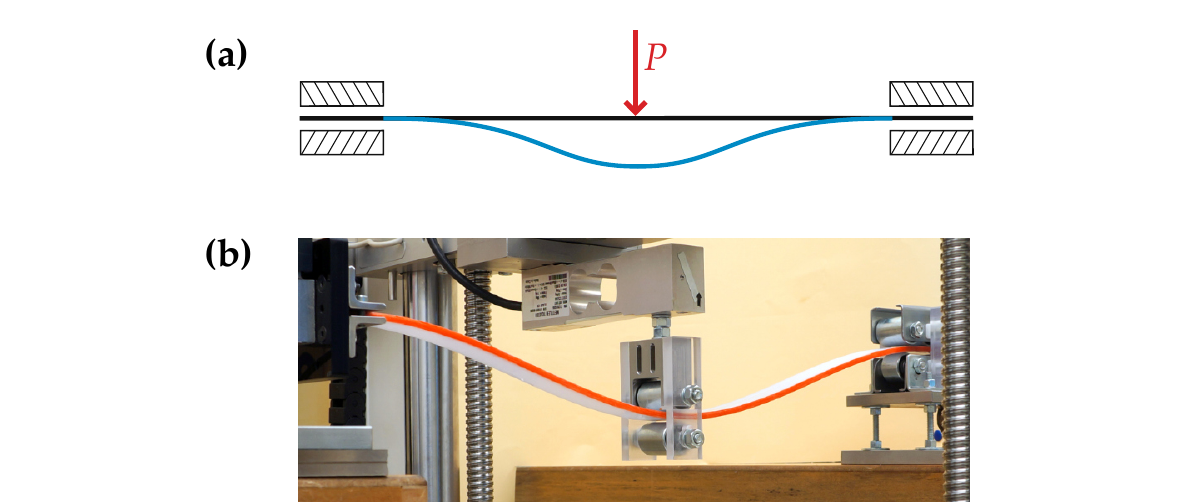}
\caption{\footnotesize (a) The encapsulation of an elastic rod, loaded with a transverse force $P$ applied at midspan, is obtained through the 
use of two sliding sleeves. The sliding sleeves provide two equal and opposite configurational forces, driving the encapsulation mechanism. (b) Experimental setup designed to perform experiments on self-encapsulation.
}
\label{rega_5}
 \end{center}
\end{figure}

\begin{figure}[H]
  \begin{center}
 \includegraphics[width=\textwidth]{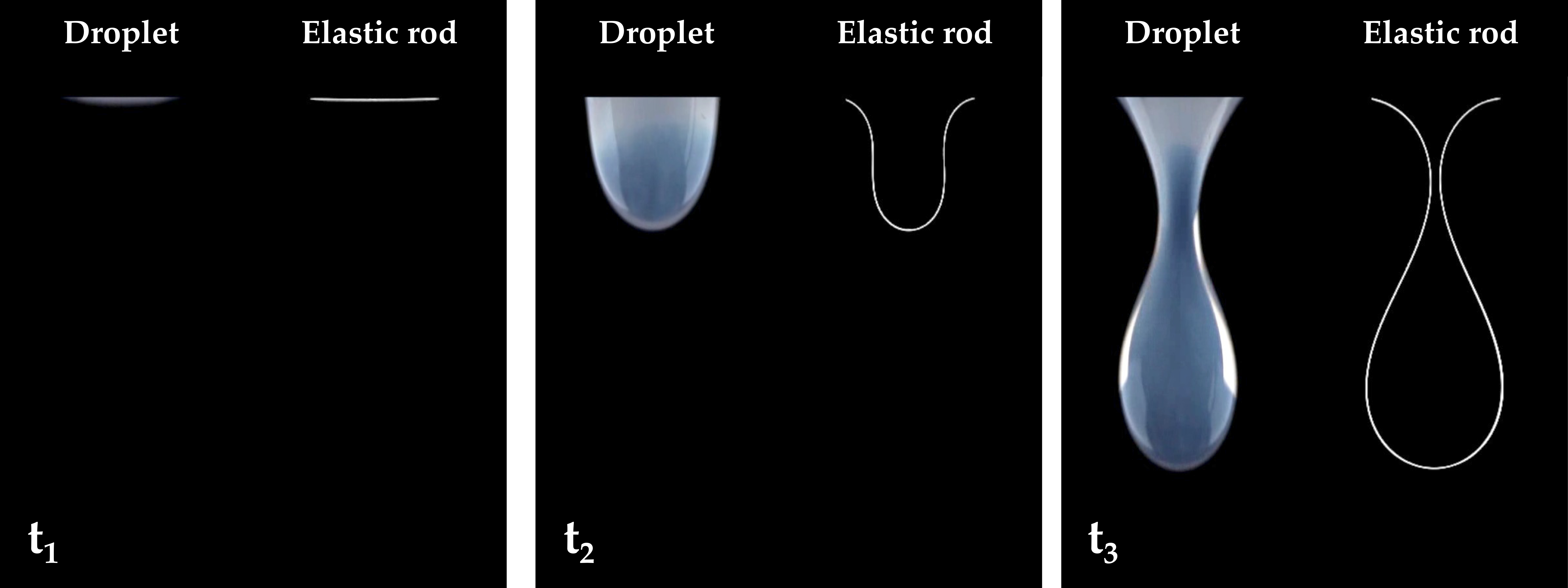}
\caption{\footnotesize The dripping of an elastic rod is compared with the progressive formation of an oil drop at three different configurations. 
}
\label{trave-goccia}
 \end{center}
\end{figure}

The new idea pursued here is obtained through the use of two sliding sleeves, as illustrated in Fig. \ref{rega_5}.
These sleeves provide two compressive configurational forces, which tend to \lq close' the rod. 
Since the differential equation of the elastica not only governs the oscillation of
a simple pendulum and the deflection of an elastic rod, but also the shape of a droplet, it can be appreciated that the 
process of encapsulation will be similar to the formation of a droplet, from which the idea of \lq dripping of an elastic rod' arose. 
The encapsulation driven by Eshelby-like forces was demonstrated to be always possible (for every rod geometrical configuration) by Bosi et al. \cite{5bomidabi1}. The dripping of an elastic rod is shown in 
Fig. \ref{trave-goccia}, where it is compared with the process of formation of an oil droplet. 
%%%%%%%%%%%%%%%%%%%%%%%%%%%%%%%%%%%%%%%%%%%%%%%%%%%%%%%%
%%%%%%%%%%%%%%%%%%%%%%%%%%%%%%%%%%%%%%%%%%%%%%%%%%%%%%%%%

%%%%%%%%%%%%%%%%%%%%%%%%%%%%%%%%%%%%%%%%%%%%%%%%%%%%%%%%
%%%%%%%%%%%%%%%%%%%%%%%%%%%%%%%%%%%%%%%%%%%%%%%%%%%%%%%%%

\section{Penetrating blades}

\begin{figure}[ht]
  \begin{center}
 \includegraphics[width=.99\textwidth]{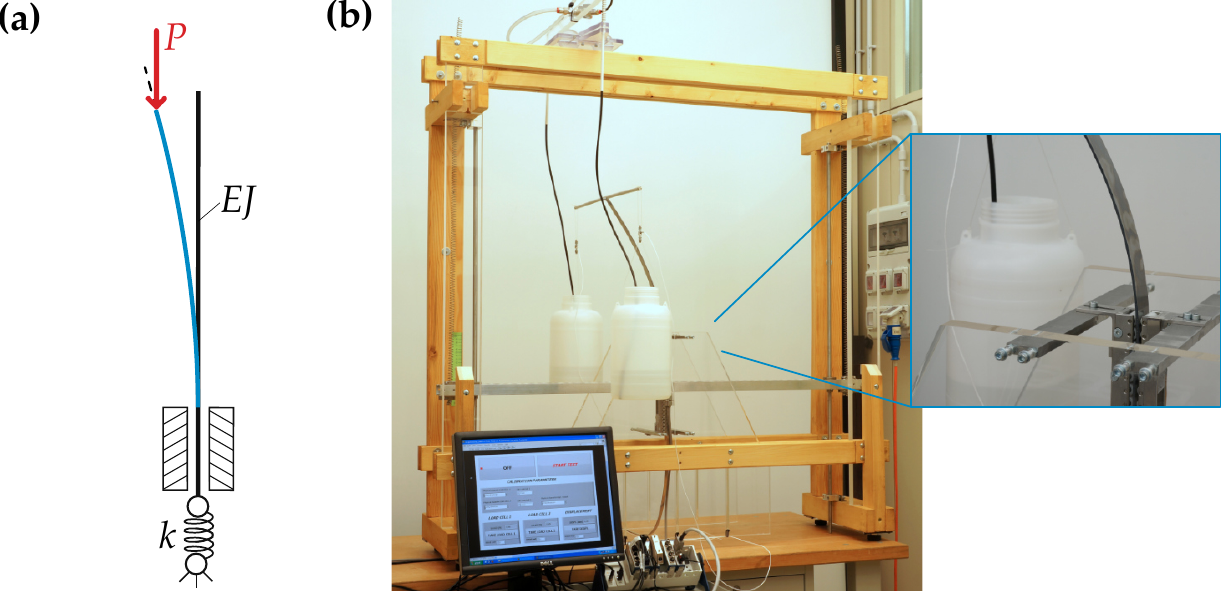}
\caption{\footnotesize Instability of a penetrating blade: the  elastic rod is loaded axially with a vertical dead force $P$ and it can slide inside a sliding sleeve placed at the  other end, while being restrained by a linear spring of stiffness $k$. Three important features emerge from this problem: (i.) an increase in the stiffness of the spring leads to a decrease of the buckling load; (ii.) the straight configuration is stable for loads smaller than the buckling load, but can return stable at higher loads; (iii.) during the post-critical behaviour a vertical upward configurational force develops, which could even eject the rod from the sleeve. (a) Schematic of the penetrating blade problem. (b) Experimental setup realized for the experimental validation of the features associated with the mechanics of the system.
}
\label{rega_4}
 \end{center}
\end{figure}

Configurational forces deeply influence the stability of structures. 
A paradigmatic example has been presented by Bigoni et al. \cite{4bubodami} and is sketched in Fig. \ref{rega_4}, where 
an elastic rod is loaded with a dead vertical force $P$ applied at one end and it can slide inside a sliding sleeve present at the other end, while being axially restrained by a linear spring of stiffness $k$. 
%%%%%%%%%%%%%%%%%%%%%%%%%%%%%%%%%%%%%%%%%%%%%%%%%%%%%%%%
%%%%%%%%%%%%%%%%%%%%%%%%%%%%%%%%%%%%%%%%%%%%%%%%%%%%%%%%%

This structure is a generalization of that reported in \cite{bigoni}, used to show two counter-intuitive effects: (i.) an increase in the stiffness $k$ of the restraining 
spring {\it lowers} the buckling load and (ii.) the straight configuration of the elastic rod may return to be stable at an axial load higher than that triggering buckling. 

During the post-critical deformation of the rod sketched in Fig. \ref{rega_4}, a configurational force progressively develops. This force is vertical and directed upwards, so that it lifts the rod and sometimes it can become so large that the rod is expelled from the sliding sleeve. 
The configurational force also plays an important role in the restabilization problem, when the straight configuration spontaneously returns stable after bifurcation, 
Bosi et al. \cite{7bomidanebi}, or when a rod is injected inside a constraint \cite{6bomidabi2}. 
Further problems of instability involving configurational forces were analyzed by 
Liakou et al. \cite{liakou1,liakou2,liakou3}.

\section{Torsional actuator}

The same concept ruling the development of a configurational force in a rod under bending can be applied under  a state of torsion. In this case, Bigoni et al. 
\cite{3bidamibo} have proven that the configurational force is given by
\beq
\frac{M^2}{2D},
\eeq
where $M$ is the twisting moment and $D$ is the torsional rigidity of the rod. 

\begin{figure}[ht]
  \begin{center}
 \includegraphics[width=1\textwidth]{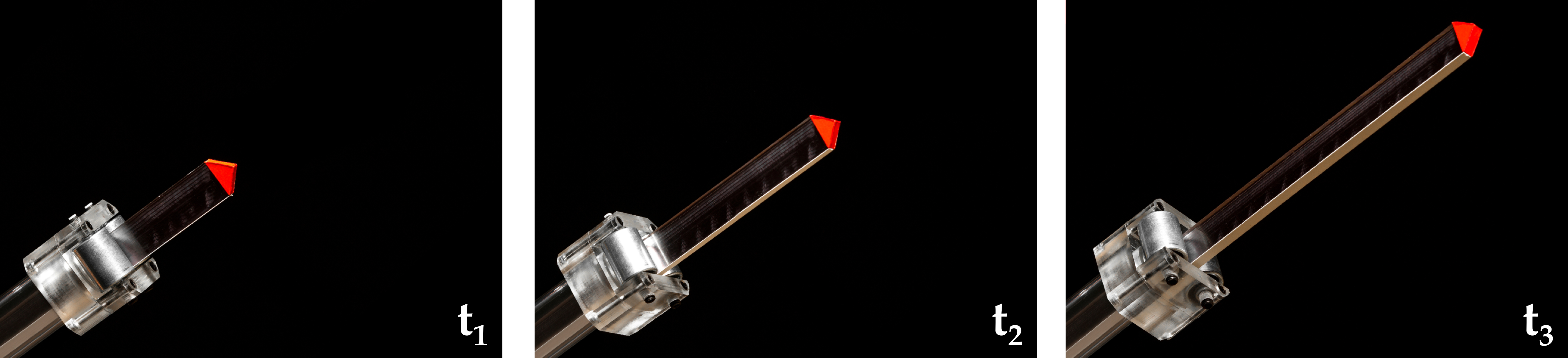}
\caption{\footnotesize The torsional actuator, based on the development of a configuration force, transforms a twist into a longitudinal trust through a release of the elastic energy, 
without any use of gears or other transmission mechanisms.
}
\label{gun}
 \end{center}
\end{figure}

This configurational force may be used to create an actuator, turning a twist into a longitudinal motion, as sketched in Fig. \ref{gun}. 
%%%%%%%%%%%%%%%%%%%%%%%%%%%%%%%%%%%%%%%%%%%%%%%%%%%%%%%%
%%%%%%%%%%%%%%%%%%%%%%%%%%%%%%%%%%%%%%%%%%%%%%%%%%%%%%%%%
The actuator represents a purely \lq elastic machine' or \lq soft device', in which the longitudinal trust is obtained without the use of gears or other transmission mechanisms.

\section{Snaking of an elastic rod}

In the problem of \lq snaking', an elastic rod, straight when unloaded, is inside a curved channel. The channel is perfectly 
frictionless, rigid and tight to the rod. In this situation, a propulsion force is developed from a release of elastic flexural energy and the rod's movement is realized. Such a  propulsion force is a configurational force, which was theoretically estimated and experimentally demonstrated by Dal Corso et al. 
\cite{9damipumomobi}. 
The obtained propulsion force has been suggested to represent the key in the limbless locomotion, typical of serpentine motion \cite{gray,kuz,nose,cicco}.

\section{Conclusions}
The presence of configurational forces on structures   was discovered for the first time, analyzed and developed by the authors of this article 
in the last decade. 
These forces have been theoretically demonstrated, experimentally validated and implemented in the realisation of novel reconfigurable elastic devices. They have been shown to strongly 
affect the behaviour of elastic structures and led to the discovery of several new effects. 
These  results 
inspire new interpretations of configurational forces in solids 
\cite{royer}
and 
pave the way to the new territory of configurational structural mechanics, towards innovative applications of reconfigurable mechanisms in several fields ranging from nanomedicine to aerospace. 

\vspace{0.5cm}
\noindent \textbf{Acknowledges.} Financial support is acknowledged from H2020- MSCA-ITN-2020-LIGHTEN-956547.

%
% ---- Bibliography ----
%
% BibTeX users should specify bibliography style 'splncs04'.
% References will then be sorted and formatted in the correct style.
%
% \bibliographystyle{splncs04}
% \bibliography{mybibliography}

\begin{thebibliography}{8}

\bibitem{eshelby1} Eshelby, J.D.: The force on an elastic singularity. Phil. Trans. Roy. Soc.  A, \textbf{244}, 87-112 (1951) 

\bibitem{eshelby2} Eshelby, J.D.: The continuum theory of lattice defects. In: Progress in Solid State Physics 3 (eds. F. Seitz and D. Turnbull) 79-144, Academic Press, New York (1956).

\bibitem{eshelby3} Eshelby, J.D.: Energy relations and the energy-momentum tensor in continuum mechanics. In: Inelastic Behaviour of Solids (eds. M. Kanninien, W. Adler, A. Rosenfield, and R. Jaffee), 77-115, McGraw-Hill, New York (1970) 

\bibitem{eshelby4} Eshelby, J.D.: The elastic energy-momentum tensor. J.  Elas.,  \textbf{5}, 321-335 (1975) 

\bibitem{gurtin} Gurtin, M.E.: Configurational Forces as Basic Concept of Continuum Physics. Springer, Berlin, New York, Heidelberg (2000) 

\bibitem{kienzler} Kienzler, R., Herrmann, G.: Mechanics in Material Space. Springer, New York, Berlin, Heidelberg (2000)

\bibitem{maugin1}  Maugin, G.A.: Material Inhomogeneities in Elasticity, Chapman and Hall, London (1993) 

\bibitem{maugin2}  Maugin G.A.: Configurational Forces: Thermodynamics, Physics, Mathematics and Numerics, Chapman \& Hall, CRC -Taylor and Francis, New York (2011) 

\bibitem{dascalu}  Dascalu, C., Maugin, G.A., Stolz, C.: Defect and Material Mechanics, Springer (2010) 

\bibitem{bigoni-deseri}  Bigoni, D., Deseri, L.: Recent Progress in the Mechanics of Defects, Springer (2011) 

\bibitem{2bodabomi} Bigoni, D., Dal Corso, F., Bosi, F., Misseroni, D.: Eshelby-like forces acting on elastic structures: theoretical and experimental proof. Mech.  Mat.,  \textbf{80}, 368--374 (2015) 

\bibitem{1armanini} Armanini, C., Dal Corso, F., Misseroni, D., Bigoni, D.:
Configurational forces and nonlinear structural dynamics. J. Mech. Phys. Sol., \textbf{130}, 82--100 (2019)

\bibitem{10datamomobi} Dal Corso, F., Tallarico, D., Movchan, N., Movchan, A., Bigoni, D.: Nested Bloch waves in elastic structures with configurational forces. Phil. Trans. Roy. Soc. A, \textbf{377}: 20190101 (2019)

\bibitem{8bodamibi} Bosi, F.,  Dal Corso, F., Misseroni, D., Bigoni, D.: An elastica arm scale. Proc. Roy. Soc. A, \textbf{470}, 20140232 (2014)

\bibitem{or1} O'Reilly, O.M.:  Some perspectives on Eshelby-like forces in the elastica arm scale. Proc. Roy. Soc. A, \textbf{471}, 20140785 (2015) 

\bibitem{or2}  O'Reilly, O.M.: Modeling Nonlinear Problems in the Mechanics of Strings and Rods: The Role of the Balance Laws, Springer (2018) 

\bibitem{vella}  Vella, D., du Pontavice, E., Hall, C.L., Goriely, A.: The magneto-elastica: from self-buckling to self-assembly. Proc. Roy. Soc. A, \textbf{470}, 20130609 (2014)  


\bibitem{shim}  Shim, J., Perdigou, C., Chen, E.R., Bertoldi, K., Reis, P.M.:  Buckling-induced encapsulation of structured elastic shells under pressure. Proc. Natl Acad. Sci. USA, \textbf{109}, 5978-5983 (2012)

\bibitem{antk}   Antkowiak, A., Audoly, B., Josserand, C., Neukirch, S., Rivetti, M.:  Instant fabrication and selection of folded structures using drop impact. Proc. Natl Acad. Sci. USA, \textbf{108}, 10401--10404 (2011)


\bibitem{rivetti}  Rivetti, M., Neukirch, S.: Instabilities in a drop-strip system: a simplified model. Proc. Roy. Soc. A, \textbf{468}, 1304-1324 (2012) 


\bibitem{5bomidabi1} Bosi, F., Misseroni, D., Dal Corso, F., Bigoni, D.: Self-encapsulation, or the \lq dripping' of an elastic rod. Proc. Roy. Soc. A, \textbf{471}: 20150195 (2015)

\bibitem{4bubodami} Bigoni, D., Dal Corso, F., Bosi, F., Misseroni, D.: Instability of a penetrating blade. J. Mech. Phys. Sol., \textbf{64}, 411-425 (2014) 


\bibitem{bigoni} Bigoni, D.: Nonlinear Solid Mechanics. Cambridge University Press (2012) 

\bibitem{7bomidanebi} Bosi, F., Misseroni, D., Dal Corso, F., Neukirch, S., Bigoni, D.: Asymptotic self-restabilization of a continuous elastic structure. Phys. Rev. E, \textbf{94}, 063005 (2016)

\bibitem{6bomidabi2} Bosi, F., Misseroni, D., Dal Corso, F., Bigoni, D.: Development of configurational forces during the injection of an elastic rod. Ext.  Mech. Lett., \textbf{4}, 83-88 (2015)



\bibitem{liakou1}
Liakou, A.: Constrained buckling of spatial elastica: Application of optimal control
method. J. App. Mech. ASME, \textbf{85}, 081005 (2018) 

\bibitem{liakou2}
Liakou, A.: Application of optimal control method in buckling analysis of constrained elastica problems. Int. J. Sol. Struct., \textbf{141-142}, 158-172 (2018) 

\bibitem{liakou3}
Liakou, A., Detournay, E.:  Constrained buckling of variable length elastica: Solution
by geometrical segmentation. Int. J. Non-Linear Mech., \textbf{99}, 204-217  (2018)



\bibitem{3bidamibo} Bigoni, D., Dal Corso, F., Misseroni, D., Bosi, F.: Torsional locomotion. Proc. Roy. Soc. A, \textbf{470} 20140599 (2014)

\bibitem{9damipumomobi} Dal Corso, F., Misseroni, D., Pugno, N.M., Movchan, A.B., Movchan, N.V., Bigoni, D.: Serpentine locomotion through elastic energy release. J. Roy. Soc. Interface, \textbf{14} 20170055 (2017)  

 
\bibitem{gray} Gray, J: The mechanism of locomotion in snakes. J. Exp. Biol., \textbf{23}, 101-120 (1946)

\bibitem{kuz} Kuznetsov, V.M., Lugovtsov, B.A., Sher, Y.N.: On the motive mechanism of snakes and fish. Arch.  Rat.  Mech.   Anal., \textbf{25}, 367–387 (1967)

\bibitem{nose} Noselli, G., DeSimone, A.: A robotic crawler exploiting directional frictional interactions: Experiments, numerics and derivation of a reduced model. Proc. Roy. Soc. A, \textbf{470} (2014)

\bibitem{cicco} Cicconofri, G., DeSimone, A.: A study of snake-like locomotion through the analysis of a flexible robot model. Proc. Roy. Soc. A, \textbf{471} 20150054 (2015)

\bibitem{royer}
Ballarini, R, Royer-Carfagni, G.:  A Newtonian interpretation of configurational
forces on dislocations and cracks. J. Mech. Phys. Sol., \textbf{95}, 602-620 (2016)





%\bibitem{ref_article1}
%Author, F.: Article title. Journal \textbf{2}(5), 99--110 (2016)

%\bibitem{ref_lncs1}
%Author, F., Author, S.: Title of a proceedings paper. In: Editor,
%F., Editor, S. (eds.) CONFERENCE 2016, LNCS, vol. 9999, pp. 1--13.
%Springer, Heidelberg (2016). \doi{10.10007/1234567890}

%\bibitem{ref_book1}
%Author, F., Author, S., Author, T.: Book title. 2nd edn. Publisher,
%Location (1999)

%\bibitem{ref_proc1}
%Author, A.-B.: Contribution title. In: 9th International Proceedings
%on Proceedings, pp. 1--2. Publisher, Location (2010)

%\bibitem{ref_url1}
%LNCS Homepage, \url{http://www.springer.com/lncs}. Last accessed 4
%Oct 2017
\end{thebibliography}
%

\end{document}